\documentclass[showpacs,aps,graphicx,twocolumn]{revtex4}

\usepackage{amsmath}
\usepackage{graphicx}

\begin{document}


\title{Universal hyperparallel hybrid photonic quantum gates with dipole-induced transparency
in the weak-coupling regime\footnote{Published in Phys. Rev. A
\textbf{91}, 032328 (2015)}}
\author{Bao-Cang Ren, Guan-Yu Wang, and Fu-Guo Deng\footnote{Corresponding author: fgdeng@bnu.edu.cn} }
\address{
Department of Physics, Applied Optics Beijing Area Major Laboratory,
Beijing Normal University, Beijing 100875, China }
\date{\today }

\begin{abstract}

We present the dipole induced transparency (DIT) of a diamond
nitrogen-vacancy center embedded in a photonic crystal cavity
coupled to two waveguides, and it is obvious with the robust and
flexible reflectance and transmittance difference of circularly
polarized lights between the uncoupled and the coupled cavities even
in the  bad cavity regime (the Purcell regime). With this DIT, we
propose two universal hyperparallel hybrid photonic quantum logic
gates, including a hybrid hyper-controlled-not gate and
a hybrid hyper-Toffoli gate, on photon systems in both the
polarization and the spatial-mode degrees of freedom (DOFs), which
are equal to two identical quantum logic gates operating
simultaneously on the systems in one DOF. They can be used to perform more
quantum operations with less resources in the quantum
information protocols with multi-qubit systems in several
DOFs, which may depress the resources consumed and the photonic dissipation.
Moreover, they are more robust against asymmetric environment noise in the
weak-coupling regime, compared with the integration of two cascaded
quantum logic gates in one DOF.


\end{abstract}
 \maketitle



\section{Introduction}

The quantum computer is powerful in quantum information processing
because of its fascinating capability of parallel computing,
according to quantum mechanics theory  \cite{QC}. Quantum logic
gates are the key elements to precisely control and manipulate
quantum states in quantum computation. Many proposals have been
proposed to implement quantum logic gates with several physical
systems both in theory and in experiment \cite{CQ}, such as the ion trap
\cite{NTI}, nuclear magnetic resonance \cite{NMR}, quantum dot
\cite{QD3}, superconducting  qubits \cite{SCCQ}, and photon systems
\cite{linear,nonlinear,mula}. In practice, there are still several
obstacles required to be overcome in the implementation of universal
quantum logic gates, especially for the interaction between qubits.
The optical nonlinearity of cavity quantum electrodynamics (QED)
holds great promise for photon-photon, photon-dipole, and
dipole-dipole interactions, and it has been used to complete some
important tasks in quantum information processing, such as
entanglement generation \cite{mula2,QD1,QD} and  quantum logic gates
\cite{QD4,QD5,mula}.

Usually, the approaches for light-dipole interaction in cavity QED
are focused on the strong-coupling regime \cite{mula,QD1,mula1},
which is always referred to the high-$Q$ regime with the vacuum Rabi
frequency of a dipole ($g$) beyond  both the cavity and the dipole
decay rates. The strong coupling between a single atom and a photon
has been demonstrated experimentally with cavity QED in the past few
years \cite{SC,SC1}, and it has been used to implement the  quantum
logic gate between a single photon and  a single trapped atom in
experiment  recently \cite{SC1}. In a bad cavity regime, called the
Purcell regime \cite{Pur} with the cavity decay rate much bigger
than the dipole decay rate, the interesting nonlinear optical
property can also be observed with a much smaller coupling strength
$g$. With the Purcell effect, the dipole-induced transparency (DIT)
can be used for quantum information processing in the weak-coupling
regime (low-$Q$ regime) \cite{QD,QD4,RT}. The fiber-optical switch
\cite{FOS} and the quantum phase switch \cite{QPS} for photons have
been demonstrated experimentally in the Purcell regime.

A nitrogen vacancy (NV) center in diamond is a promising candidate
for a solid-state matter qubit (a dipole emitter) in cavity QED due
to its long electron-spin decoherence time even at room
temperature
 \cite{NV}. The approaches about an NV center in diamond coupled to
 an optical cavity (or a nanomechanical resonator) have been
investigated both in theory  \cite{CNV} and in experiment
 \cite{CNV1,CNV3,CNV4,CNV2,CNV5}. The NV-center spins in diamonds are
very useful in quantum networks for algorithms and quantum memories
 \cite{NV7,NV8}. The quantum entanglement between the polarization of
a single photon and the electron spin of an NV center in diamond has
been produced in experiment  \cite{NV2}, and the Faraday effect
induced by the single spin of an NV center in diamond coupling to light
has  been observed in experiment as well \cite{NV3}, which have
facilitating applications in quantum information processing.

In this paper, we show that the DIT of a double-sided
cavity-NV-center system (an NV center in diamond embedded in a
photonic crystal cavity coupled to two waveguides)  can be used for
the photon-photon interaction in both the polarization and
spatial-mode degrees of freedom (DOFs). In the Purcell regime, the
DIT  is still obvious with the robust and flexible reflectance and
transmittance difference of circularly polarized lights between the
uncoupled and the coupled cavities. With this DIT, we construct a
hybrid polarization-spatial hyper-controlled-not (CNOT) gate
on a two-photon system, which is equal to two CNOT gates operating
simultaneously on a four-photon system in one DOF. Also, we present
a hybrid  polarization-spatial hyper-Toffoli gate on a three-photon
system and it is equal to two Toffoli gates on a six-photon system
in one DOF. These universal hyperparallel hybrid photonic quantum
gates can reduce the resources consumed in quantum information
processing, and they are more robust against the photonic
dissipation noise, compared with the integration of two cascaded
quantum logic gates in one DOF. They have high fidelities in the
symmetrical regime of double-sided cavity-NV-center systems and they
can depress the asymmetric environment noise in the weak-coupling
regime with a small Purcell factor. They can form universal
hyperparallel photonic quantum computing assisted by single-photon
rotations. Besides, they are useful for the quantum information
protocols with multi-qubit systems in several DOFs, for example, the
preparation of two-photon hyperentangled states and the complete
analysis for them.

%

\section{DIT for double-sided cavity-NV-center system}\label{sec2}

\begin{figure}[htbp]             
\centering\includegraphics[width=8 cm]{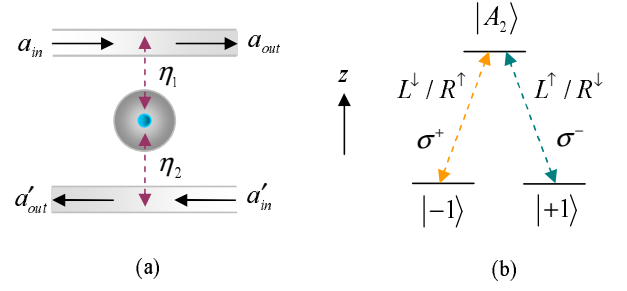} \caption{(Color
online) The optical transitions of an NV center with circularly
polarized lights. (a) A double-sided cavity-waveguide-NV-center
system. (b) The optical transitions of an NV center. The photon in
the state $|R^\uparrow\rangle$ or $|L^\downarrow\rangle$ corresponds
to $\sigma^+$, and the photon in the state $|R^\downarrow\rangle$ or
$|L^\uparrow\rangle$  corresponds  to $\sigma^-$. $R^\uparrow$
($R^\downarrow$) and $L^\uparrow$ ($L^\downarrow$) represent the
right- and left- circularly polarized lights with their input
(output) directions parallel (antiparallel) to the $z$ direction.}
\label{figure1}
\end{figure}

A negatively charged NV center in diamond consists of a
substitutional nitrogen atom, an adjacent vacancy, and six electrons
coming from the nitrogen atom and three carbon atoms surrounding the
vacancy. Its ground state is an electron-spin triplet with the
splitting at  $2.88$ GHz between the magnetic sublevels $|0\rangle$
($|m_s=0\rangle$) and $|\pm1\rangle$ ($|m_s=\pm1\rangle$). There are
six electronic excited states according to the Hamiltonian with the
spin-orbit and spin-spin interactions and $C_{3v}$ symmetry
\cite{NV4}. Optical transitions between the ground states and the
excited states are spin preserving, while the electronic orbital
angular momentum is changed by the photon polarization. The excited
state $|A_2\rangle$, which is robust with the stable symmetric
properties,  decays  with an equal probability to the ground states
$|-1\rangle$ and $|+1\rangle$ through the $\sigma^+$ and $\sigma^-$
polarization radiations, respectively  \cite{NV2} (see
Fig.\ref{figure1}(b)). The excited state $|A_2\rangle$ has the form
$|A_2\rangle=(|E_-\rangle|+1\rangle+|E_+\rangle|-1\rangle)/\sqrt{2}$
 \cite{NV2}, where $|E_\pm\rangle$ are the orbit states with the
angular momentum projections $\pm1$ along the NV axis (the $z$ axis in
Fig.\ref{figure1}). The ground states are associated with the orbit
state $|E_0\rangle$ with the angular momentum projection zero along
the NV axis.

The DIT of the cavity-NV-center system (shown in Fig.\ref{figure1}(a))
can be calculated by the Heisenberg equations of motion for the
cavity field operator $\hat{a}$ and the dipole operator
$\hat{\sigma}_-$  \cite{QD,QD11}, that is,
\begin{eqnarray}                           
\begin{split}
\frac{d\hat{a}}{dt}\!=&\!-\left[i(\omega_c-\omega)+\eta+\frac{\kappa}{2}\right]\hat{a} -\sqrt{\eta}(\hat{a}_{in}+\hat{a}'_{in}) \\
&-g\hat{\sigma}_--\hat{h}, \\
\frac{d\hat{\sigma}_-}{dt}\!=&\!-\left[i(\omega_k-\omega)+\frac{\gamma}{2}\right]\hat{\sigma}_--g\hat{\sigma}_z\hat{a}-\hat{f}.
\end{split}
\end{eqnarray}
Here, $\omega_k$ ($k=-1,+1$), $\omega$, and $\omega_c$ are the
frequencies of the transition between $|-1\rangle$ ($|+1\rangle$)
and $|A_2\rangle$, the waveguide channel mode, and the cavity mode,
respectively. $g$ is the coupling strength of the cavity to the
NV center. $\gamma/2$ is the decay rate of the emitter. $\eta$ and
$\kappa/2$ are the decay rates of the cavity field into waveguide
channel modes and cavity intrinsic loss modes, respectively.
$\hat{g}$ and $\hat{f}$ are noise operators, which can preserve
the commutation relation. The operators $\hat{a}_{in}$ ($\hat{a}'_{in}$)
and $\hat{a}_{out}$ ($\hat{a}'_{out}$) are the input and output
field operators, respectively. They satisfy the boundary relations
$\hat{a}_{out}=\hat{a}_{in}+\sqrt{\eta}\,\hat{a}$ and
$\hat{a}'_{out}=\hat{a}'_{in}+\sqrt{\eta}\,\hat{a}$.  The decay
rates of the cavity field into two waveguides can be set very close to
get approximately the same fidelity for both directions
($\eta_1\cong\eta_2=\eta$)  \cite{SC}. In the weak excitation limit
with the emitter predominantly in the ground state
($\langle\sigma_z\rangle=-1$), the transmission and reflection
coefficients of the cavity-NV-center system are given by
\begin{eqnarray}                           
\begin{split}
t(\omega)&=\frac{-\eta[i(\omega_k-\omega)+\frac{\gamma}{2}]}{[i(\omega_k-\omega)
+\frac{\gamma}{2}][i(\omega_c-\omega)+\eta+\frac{\kappa}{2}]+g^2},\\
r(\omega)&=1+t(\omega).
\end{split}
\end{eqnarray}

\begin{figure}[htb]                    
\centering
\includegraphics[width=7.2 cm]{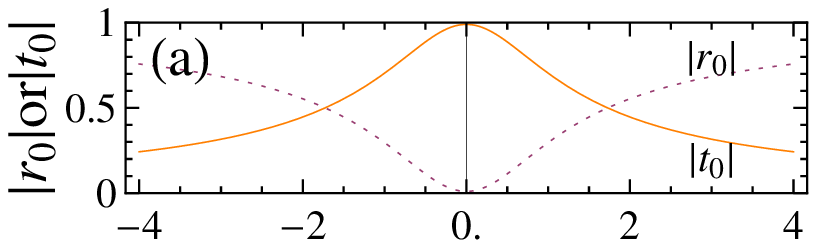}
\includegraphics[width=7.2 cm]{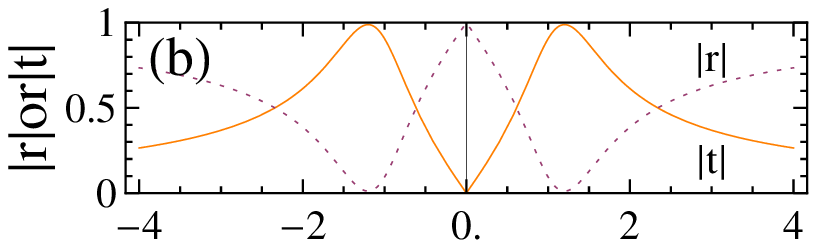}
\includegraphics[width=7.2 cm]{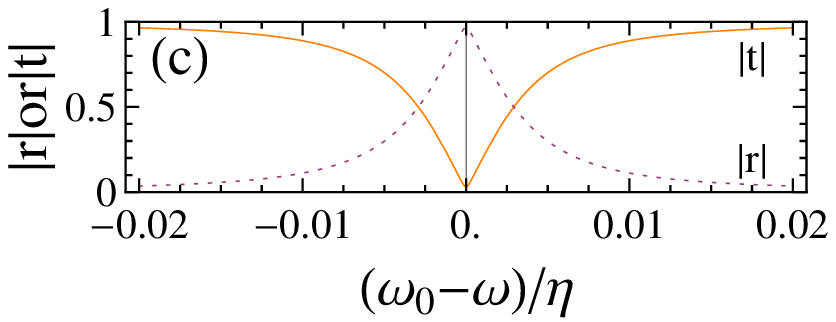}
\caption{(Color online) The reflection and transition coefficients
of the double-sided cavity-NV-center system vs the normalized
frequency detuning $(\omega-\omega_0)/\eta$
($\omega_c=\omega_k=\omega_0$). (a) $g=0$, $\gamma\sim2\pi\times80$
MHz \cite{CNV2} and $\eta=50\kappa\sim2\pi\times0.05$ THz
($Q\sim10^4$). (b) $g\sim2\pi\times0.06$ THz,
$\gamma\sim2\pi\times80$ MHz, and $\eta=50\kappa\sim2\pi\times0.05$
THz. (c) $g\sim2\pi\times0.035$ THz, $\gamma\sim2\pi\times80$ MHz,
and $\eta=50\kappa\sim2\pi\times0.5$ THz ($Q\sim10^3$).}
\label{figure2}
\end{figure}

Considering the emitter is resonant with the cavity mode
($\omega_c=\omega_k=\omega$), the reflection and transmission
coefficients are $t=-(2F_p+1+\frac{\lambda}{2})^{-1}$ and
$r=(2F_p+\frac{\lambda}{2})/(2F_p+1+\frac{\lambda}{2})$ for $g>0$,
and they are $t_0=-(1+\frac{\lambda}{2})^{-1}$ and
$r_0=\frac{\lambda}{2}/(1+\frac{\lambda}{2})$ for $g=0$. Here
$F_p=g^2/(\eta\gamma)$ is the Purcell factor ($\kappa\approx0$), and
$\lambda=\kappa/\eta$. If the Purcell factor is $F_p\gg1$,
the reflection and transmission coefficients are
$r(\omega)\rightarrow1$ and $t(\omega)\rightarrow0$. If the cavity
decay rate is $\lambda\ll1$, the reflection and transmission
coefficients of the bare cavity are $r_0(\omega)\rightarrow0$ and
$t_0(\omega)\rightarrow-1$ (Fig.\ref{figure2} (a)). The interaction
between  a single photon and the emitter in an NV center is obtained
as
\begin{eqnarray}                           
\begin{split}
|\sigma^+\rangle(|-1\rangle+|+1\rangle)\;\;\rightarrow\;\;& |\sigma^+_r\rangle|-1\rangle-|\sigma^+_{t_0}\rangle|+1\rangle,\\
|\sigma^-\rangle(|-1\rangle+|+1\rangle)\;\;\rightarrow\;\;&-|\sigma^-_{t_0}\rangle|-1\rangle+|\sigma^-_r\rangle|+1\rangle.
\label{eq3}
\end{split}
\end{eqnarray}
Here the subscript $r$ ($t_0$) represents the photon  reflected
(transmitted).

In the strong-coupling (high-$Q$) regime, the dipole-induced
reflection is the result of vacuum Rabi splitting with the Rabi
frequency $\Omega=2g$, and the transmission (reflection) dip is
equal to $2g$ (Fig.\ref{figure2}(b)). The incoming pulse must be
longer than the Rabi oscillation period $1/g$ in this high-$Q$
regime \cite{QD}. In the weak-coupling (low-$Q$) regime, the
dipole-induced reflection is caused by the destructive interference of
the cavity field and the dipole emission field, and the transmission
(reflection) dip is equal to $2\Gamma=2F_p\gamma/(1+\frac{\lambda}{2})$
(Fig.\ref{figure2}(c)). The incoming pulse must be longer than the Rabi
oscillation period $1/\Gamma$ in this bad cavity regime  \cite{QD}.
In the weak excitation approximation, the time interval between two
photons should be longer than $\Delta\tau=2F_p/[\gamma(1+\frac{\lambda}{2})]$.

\begin{figure}[tpb]        
\centering\includegraphics[width=8.0 cm,angle=0]{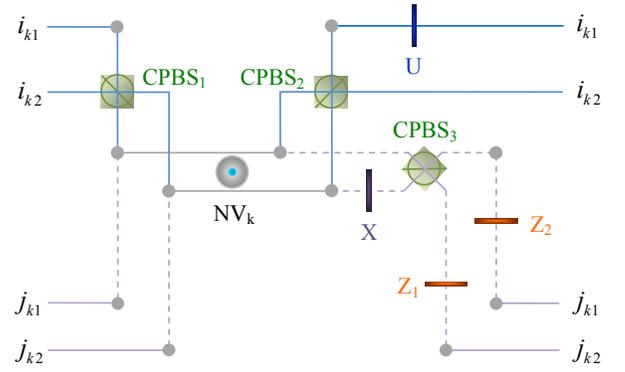}
\caption{(Color online) Schematic diagram for a hybrid photonic
hyper-CNOT gate operating on  a two-photon system in both the
spatial-mode and polarization DOFs. $X$ represents a half-wave plate
which is used to perform a polarization bit-flip operation
$X=|R\rangle\langle L|+|L\rangle\langle R|$. $Z_n$ ($n=1,2$) represents a
half-wave plate which is used to perform a polarization phase-flip
operation $Z=|R\rangle\langle R|-|L\rangle\langle L|$. $U$ represents
a wave plate which is used to perform a polarization phase-flip
operation $U=-|R\rangle\langle R|-|L\rangle\langle L|$. CPBS$_m$ ($m=1,2,3$)
represents a polarizing beam splitter in the circular basis, which
transmits the photon in right-circular polarization $\vert R\rangle$
and reflects the photon in left-circular polarization $\vert
L\rangle$, respectively. $i_{k1}$ and $i_{k2}$ represent the two
spatial modes of photon $i$ ($i=a,b$), respectively. NV$_k$
($k=1,2$) represents a double-sided cavity-NV-center system.
An optical switch is used in the merging point of $i_{kl}$
and $j_{kl}$.
\label{figure3}}
\end{figure}

The transmission and reflection rule in Eq.(\ref{eq3}) can be
described in the circular basis $\{|R\rangle, |L\rangle\}$ shown in
Fig.\ref{figure1}(b). The photon circular polarization is usually
related to the direction propagation, and the handedness circular
polarized light is changed after reflection. That is,
\begin{eqnarray}                           
\begin{split}
&|R^\uparrow,-1\rangle \;\rightarrow\;
|L^\downarrow,-1\rangle,\;\;\;\;\;\;  |R^\uparrow,+1\rangle
\;\rightarrow\;
-|R^\uparrow,+1\rangle,\\
&|L^\downarrow,-1\rangle \;\rightarrow\;
|R^\uparrow,-1\rangle,\;\;\;\;\;\;  |L^\downarrow,+1\rangle
\;\rightarrow\; -|L^\downarrow,+1\rangle,\;\;\;\;\;  \\
&|R^\downarrow,-1\rangle \;\rightarrow\;
-|R^\downarrow,-1\rangle,\;\;\;  |R^\downarrow,+1\rangle
\;\rightarrow\;
|L^\uparrow,+1\rangle,\\
&|L^\uparrow,-1\rangle \;\rightarrow\;
-|L^\uparrow,-1\rangle,\;\;\;\;  |L^\uparrow,+1\rangle
\;\rightarrow\; |R^\downarrow,+1\rangle.\label{eq4}
\end{split}
\end{eqnarray}
Here, in the left-hand side of "$\rightarrow$" in Eq.(\ref{eq4}), $|R^\uparrow\rangle$ ($|L^\uparrow\rangle$) represents that the photon
$R$ ($L$) is put into the cavity-NV-center system through  the down spatial mode of the cavity-NV-center system,
and $|R^\downarrow\rangle$ ($|L^\downarrow\rangle$) represents that the photon
$R$ ($L$) is put into the cavity-NV-center system through the upper spatial mode of the cavity-NV-center system.
In the right-hand side of "$\rightarrow$" in Eq.(\ref{eq4}), $|R^\uparrow\rangle$ ($|L^\uparrow\rangle$) represents that the photon
$R$ ($L$) exits from the cavity-NV-center system through  the upper spatial mode of the cavity-NV-center system,
and $|R^\downarrow\rangle$ ($|L^\downarrow\rangle$) represents that the photon
$R$ ($L$) exits from the cavity-NV-center system through the down spatial mode of the cavity-NV-center system.
$i_1$ and $i_2$ represent the two spatial modes of photon $i$
($i=a,b$) as shown in Fig.\ref{figure3}.

\section{Hybrid photonic hyper-CNOT gate on a two-photon system}\label{sec3}

Here, a hybrid photonic hyper-CNOT gate on  a two-photon system in
both the polarization and spatial-mode DOFs is used to complete the
task that  a bit-flip operation is performed on the spatial mode of
photon $b$ (the target qubit) when the polarization of photon $a$
(the control qubit) is in the state $\vert L\rangle$, and
simultaneously  a bit-flip operation takes place on the spatial mode
of photon $a$ when the polarization of photon $b$ is in the state
$\vert L\rangle$. It can act as two cascaded hybrid CNOT gates on a
four-photon system in one DOF with less operation time and less
resources, which is far different from the hybrid CNOT gate based on
one DOF of photon systems \cite{oneDOF}. The principle of our hybrid
photonic hyper-CNOT gate is shown in Fig.\ref{figure3}, where two
identical quantum circuits are required. We describe it in
detail as follows.

Suppose that the initial states of the two NV centers are
$|+\rangle_{e_1}$ and $|+\rangle_{e_2}$, respectively, and the
initial states of the two photons $a$ and $b$ are
\begin{eqnarray}                           
\begin{split}
|\psi_a\rangle_0&\;=\;(\alpha_1|R\rangle+\alpha_2|L\rangle)_a(\gamma_1|a_1\rangle+\gamma_2|a_2\rangle),\\
|\psi_b\rangle_0&\;=\;(\beta_1|R\rangle+\beta_2|L\rangle)_b(\delta_1|b_1\rangle+\delta_2|b_2\rangle).
\end{split}
\end{eqnarray}
Here $|\pm\rangle=\frac{1}{\sqrt{2}}(|-1\rangle\pm|+1\rangle)$.

First, we perform the Hadamard operations on the polarization DOF of
both photons $a$ and $b$, and the states of the two photons $a$ and
$b$ become
$|\psi'_a\rangle_0=(\alpha'_1|R\rangle+\alpha'_2|L\rangle)_a(\gamma_1|a_1\rangle+\gamma_2|a_2\rangle)$
and $
|\psi'_b\rangle_0=(\beta'_1|R\rangle+\beta'_2|L\rangle)_b(\delta_1|b_1\rangle+\delta_2|b_2\rangle)$.
 Here, $\alpha'_1=\frac{1}{\sqrt{2}}(\alpha_1+\alpha_2)$,
$\alpha'_2=\frac{1}{\sqrt{2}}(\alpha_1-\alpha_2)$,
$\beta'_1=\frac{1}{\sqrt{2}}(\beta_1+\beta_2)$, and
$\beta'_2=\frac{1}{\sqrt{2}}(\beta_1-\beta_2)$. The Hadamard
operation  on the polarization DOF of a photon is used to implement
the unitary single-qubit operation $\vert R\rangle \rightarrow
\frac{1}{\sqrt{2}}(\vert R\rangle + \vert L\rangle)$ and $\vert
L\rangle \rightarrow \frac{1}{\sqrt{2}}(\vert R\rangle - \vert
L\rangle)$.  Subsequently,  we lead the two  wavepackets of photon
$a$ ($b$) from the two spatial modes $|a_1\rangle$ ($|b_1\rangle$)
and $|a_2\rangle$ ($|b_2\rangle$) to spatial ports $i_{11}$
($i_{21}$) and $i_{12}$ ($i_{22}$) of the cavity-NV-center system
NV$_1$ (NV$_2$) as shown in Fig.\ref{figure3}. After photon $a$
($b$) passes through CPBS$_1$, NV$_1$ (NV$_2$), CPBS$_2$, and $U$, the state
of the quantum system composed of photon $a$ ($b$) and NV$_1$
(NV$_2$) is transformed from
$|\Psi'_{ae_1}\rangle_0\equiv|\psi'_a\rangle_0\otimes|+\rangle_{e_1}$
($|\Psi'_{be_2}\rangle_0\equiv|\psi'_b\rangle_0\otimes|+\rangle_{e_2}$)
to $|\Psi_{ae_1}\rangle_1$ ($|\Psi_{be_2}\rangle_1$). Here
\begin{eqnarray}                           
\begin{split}
|\Psi_{ae_1}\rangle_1\;=\;&\frac{1}{\sqrt{2}}\{\gamma_1[|-1\rangle_{e_1}(\alpha'_1|R\rangle+\alpha'_2|L\rangle)_a
\\
&-|+1\rangle_{e_1}(\alpha'_2|R\rangle+\alpha'_1|L\rangle)_a]|a_1\rangle \\
&+\gamma_2[|-1\rangle_{e_1}(\alpha'_2|R\rangle+\alpha'_1|L\rangle)_a \\
&-|+1\rangle_{e_1}(\alpha'_1|R\rangle+\alpha'_2|L\rangle)_a]|a_2\rangle\},\\
|\Psi_{be_2}\rangle_1\;=\;&\frac{1}{\sqrt{2}}\{\delta_1[|-1\rangle_{e_2}(\beta'_1|R\rangle+\beta'_2|L\rangle)_b
\\
&-|+1\rangle_{e_2}(\beta'_2|R\rangle+\beta'_1|L\rangle)_b]|b_1\rangle \\
&+\delta_2[|-1\rangle_{e_2}(\beta'_2|R\rangle+\beta'_1|L\rangle)_b \\
&-|+1\rangle_{e_2}(\beta'_1|R\rangle+\beta'_2|L\rangle)_b]|b_2\rangle\}.
\end{split}
\end{eqnarray}

Second, after a Hadamard operation is performed on NV$_1$ (NV$_2$),
we let photon $a$ ($b$) pass through two spatial paths $j_{21}$
($j_{11}$) and $j_{22}$ ($j_{12}$) of the cavity-NV-center system
NV$_2$ (NV$_1$) shown in Fig.\ref{figure3} (with optical switches).
Here a Hadamard operation on an NV center is used to complete the
transformations $\vert -1\rangle \rightarrow \vert +\rangle$ and
$\vert +1\rangle  \rightarrow  \vert -\rangle$. After photon $a$
($b$) passes through NV$_2$ (NV$_1$), $X$, CPBS$_3$, $Z_1$, and $Z_2$, the state of
the quantum system composed of photons $a$ and $b$, NV$_1$, and
NV$_2$ is changed from
$|\Psi_{abe_1e_2}\rangle_1\equiv|\Psi_{ae_1}\rangle_1\otimes|\Psi_{be_2}\rangle_1$
to
\begin{eqnarray}                           
|\Psi_{abe_1e_2}\rangle_2\!\!&=&\!\!\frac{1}{2}[|-1\rangle_{e_1}\alpha_2(|L\rangle-|R\rangle)_a(\delta_2|b_1\rangle+\delta_1|b_2\rangle)\nonumber\\
&&-|+1\rangle_{e_1}\alpha_1(|R\rangle+|L\rangle)_a(\delta_1|b_1\rangle+\delta_2|b_2\rangle)]\nonumber\\
&&\otimes[|-1\rangle_{e_2}\beta_2(|L\rangle-|R\rangle)_b(\gamma_2|a_1\rangle+\gamma_1|a_2\rangle)\nonumber\\
&&-|+1\rangle_{e_2}\beta_1(|R\rangle+|L\rangle)_b(\gamma_1|a_1\rangle+\gamma_2|a_2\rangle)].\nonumber\\
\end{eqnarray}

At last, with the Hadamard operations performed on NV$_1$, NV$_2$, and the
polarization DOF of photons $a$ and $b$ again,  the outcome of a
hybrid photonic hyper-CNOT gate can be obtained by measuring the two
NV centers in the orthogonal basis $\{|-1\rangle,|+1\rangle\}$ and
performing conditional phase shift operations on the polarization
modes of photons $a$ and $b$. After we perform an additional
sign change $|L\rangle_a\rightarrow-|L\rangle_a$ on photon $a$
when NV$_1$ is in the state $|+1\rangle_{e_1}$ and an addition sign change
$|L\rangle_b\rightarrow-|L\rangle_b$ on photon $b$ when NV$_2$ is in the
state $|+1\rangle_{e_2}$,  the state of the two-photon system $ab$
becomes
\begin{eqnarray}                           \label{eq.7}  
|\psi_{ab}\rangle \!\!&=&\!\!
[\alpha_1|R\rangle_a(\delta_1|b_1\rangle \!+\! \delta_2|b_2\rangle)
\!+\! \alpha_2|L\rangle_a(\delta_2|b_1\rangle\!+\!\delta_1|b_2\rangle) ]\nonumber\\
&&\!\!\otimes
[\beta_1|R\rangle_b(\gamma_1|a_1\!\rangle\!+\!\gamma_2|a_2\!\rangle)\!+\!\beta_2|L\!\rangle_b(\gamma_2|a_1\!\rangle\!+\!\gamma_1|a_2\!\rangle)].\nonumber\\
\end{eqnarray}
It is the result of a hybrid photonic hyper-CNOT gate operating
on a two-photon system, by using the polarization mode of one photon
as the control qubit and the spatial mode of the other photon as the
target qubit, respectively.

\begin{figure*}[htb]                    
\centering
\includegraphics[width=12 cm]{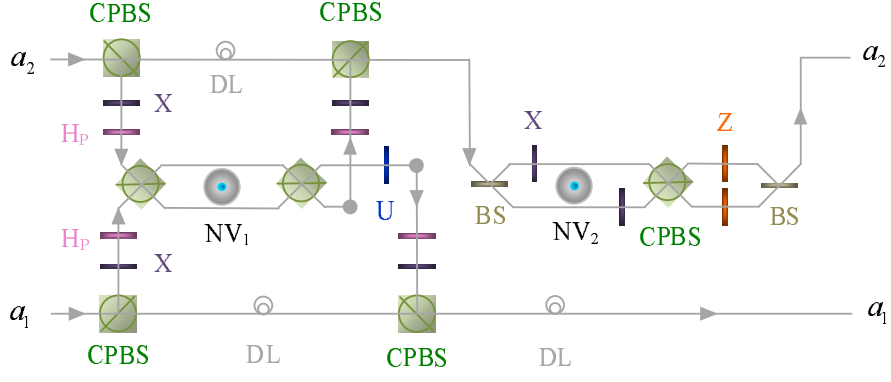}
\caption{(Color online) Schematic diagram for the first step of our
hybrid photonic hyper-Toffoli gate operating on both the
spatial-mode and polarization DOFs of a three-photon system. $H_P$
represents a  half-wave  plate which is used to perform a
polarization Hadamard operation on a photon. BS
represents a 50:50 beam splitter which is used to perform a
spatial-mode Hadamard operation on a photon
$[|a_2\rangle\rightarrow\frac{1}{\sqrt{2}}(|a'_1\rangle+|a'_2\rangle)]$.
DL represents a time-delay device. } \label{figure4}
\end{figure*}

\section{Hybrid photonic hyper-Toffoli gate on a three-photon system}\label{sec4}

A Toffoli gate is used to complete a bit-flip operation on the state of
the target qubit when both two control qubits are in the state
$\vert 1\rangle$; otherwise, nothing is done on the target qubit
\cite{QC}. It is a universal quantum gate for quantum computing.
Here, the hybrid hyper-Toffoli gate, operating on a three-photon
system $abc$ in both the polarization and spatial-mode DOFs, is used
to achieve the task that a bit-flip operation is performed on the
spatial mode of photon $c$ (the target qubit) when the polarizations
of both  photons $a$ and $b$ (the control qubits) are $\vert
L\rangle$, and simultaneously a bit-flip operation takes place on
the spatial mode of photon $b$ (the target qubit) when the
spatial mode of photon $a$ is $\vert a_2\rangle$ and the
polarization of photon $c$ is $\vert L\rangle$ (the control qubits).

\begin{figure*}[htb]                    
\centering
\includegraphics[width=13.2 cm]{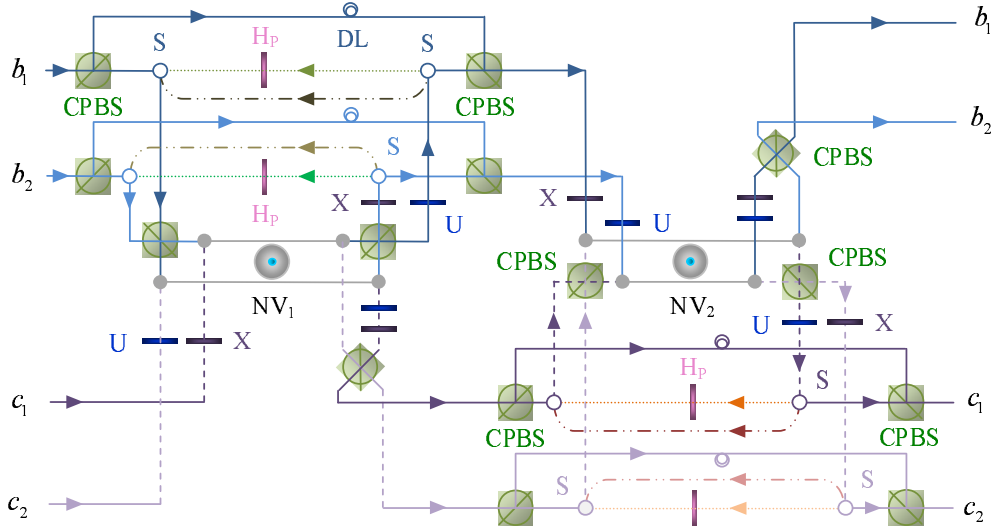}
\caption{(Color online) Schematic diagram for the second step of our
hybrid photonic hyper-Toffoli gate operating on both the
spatial-mode and polarization DOFs of a three-photon system. $S$ represents an optical switch.} \label{figure5}
\end{figure*}

The two parts of the quantum circuit for our hybrid photonic
hyper-Toffoli gate are shown in Fig.\ref{figure4} and
Fig.\ref{figure5}, respectively. Suppose that the initial states of
the two NV centers are $|+\rangle_{e_1}$ and $|+\rangle_{e_2}$,
respectively, and the initial states of three photons $a$, $b$, and
$c$ are
$|\phi_a\rangle_0=(\alpha_1|R\rangle+\beta_1|L\rangle)_a(\gamma_1|a_1\rangle+\delta_1|a_2\rangle)$,
$|\phi_b\rangle_0=(\alpha_2|R\rangle+\beta_2|L\rangle)_b(\gamma_2|b_1\rangle+\delta_2|b_2\rangle)$,
and
$|\phi_c\rangle_0=(\alpha_3|R\rangle+\beta_3|L\rangle)_c(\gamma_3|c_1\rangle+\delta_3|c_2\rangle)$,
respectively. This hyper-Toffoli gate can be constructed with two
steps described in detail below.

The principle of the first step for our hybrid photonic
hyper-Toffoli gate is shown in Fig.\ref{figure4}. First, the
wave packets of photon $a$ from the two spatial modes $|a_1\rangle$
and $|a_2\rangle$ are led to CPBS, $X$, $H_P$, CPBS, NV$_1$, CPBS, $U$, $H_P$,
$X$, and CPBS in sequence as shown in Fig.\ref{figure4}, and the
state of the quantum system composed of photon $a$ and NV$_1$ is
transformed from $|\Phi_{ae_1}\rangle_0\equiv
|\phi_a\rangle_0\otimes|+\rangle_{e_1}$ to
\begin{eqnarray}                           
|\Phi_{ae_1}\rangle_1\!\!&=&\!\!(\alpha_1|R\rangle_a|+\rangle_{e_1}+\beta_1|L\rangle_a|-\rangle_{e_1})(\gamma_1|a_1\rangle+\delta_1|a_2\rangle).\nonumber\\
\end{eqnarray}
Subsequently,  we lead the wave packet of photon $a$ from the spatial
mode $|a_2\rangle$ to BS, $X$, NV$_2$, $X$, CPBS, $Z$, and BS in sequence
as shown in Fig.\ref{figure4}, and the state of the quantum system
composed of photon $a$, NV$_1$, and NV$_2$ is transformed from
$|\Phi_{ae_1e_2}\rangle_1\equiv|\Phi_{ae_1}\rangle_1\otimes|+\rangle_{e_2}$
to
\begin{eqnarray}                           
\begin{split}
|\Phi_{ae_1e_2}\rangle_2\;=\;&(\alpha_1|R\rangle_a|+\rangle_{e_1}+\beta_1|L\rangle_a|-\rangle_{e_1}) \\
&\otimes(\gamma_1|a_1\rangle|+\rangle_{e_2}+\delta_1|a_2\rangle|-\rangle_{e_2}).
\end{split}
\end{eqnarray}

In the second step, after a Hadamard operation is performed on each
of NV$_1$ and NV$_2$,  the two wave packets of photon $b$ from the two
spatial modes $|b_1\rangle$ and $|b_2\rangle$ are led to CPBSs,
NV$_1$, CPBS, $X$, and $U$ in sequence as shown in Fig.\ref{figure5},
and the state of the quantum system composed of NV$_1$, NV$_2$,
and photons $a$ and $b$ is transformed from $|\Phi_{abe_1e_2}\rangle_2=|\Phi_{ae_1e_2}\rangle_2\otimes|\phi_b\rangle_0$  to 
\begin{eqnarray}                           
|\Phi_{abe_1e_2}\rangle_3\!\!&=&\!\!(\alpha_1\beta_2|-1\rangle_{e_1}|RL\rangle_{ab}+\beta_1\beta_2|+1\rangle_{e_1}|LL\rangle_{ab}\nonumber\\
&&\!\!-\alpha_1\alpha_2|-1\rangle_{e_1}|RL\rangle_{ab}+\beta_1\alpha_2|+1\rangle_{e_1}|LR\rangle_{ab})\nonumber\\
&&\!\!\otimes(\gamma_1|a_1\rangle|-1\rangle_{e_2}+\delta_1|a_2\rangle|+1\rangle_{e_2})\nonumber\\
&&\!\!\otimes(\gamma_2|b_1\rangle+\delta_2|b_2\rangle).
\end{eqnarray}
After a Hadamard operation is performed on NV$_1$, the two
wave packets of photon $b$ from the two spatial modes $|b_1\rangle$
and $|b_2\rangle$ are led to $H_P$, CPBS, NV$_1$, CPBS, $X$, and $U$ in
sequence with an optical switch $S$ (through the dotted line in
Fig.\ref{figure5}), and then the state of the quantum system
composed of NV$_1$, NV$_2$, and photons $a$ and $b$ becomes
\begin{eqnarray}                           
|\Phi_{abe_1e_2}\rangle_4\!\!&=&\!\![\alpha_1\beta_2|+\rangle_{e_1}|RL\rangle_{ab}+\beta_1\beta_2|-\rangle_{e_1}|LL\rangle_{ab}\nonumber\\
&&-\frac{\alpha_1\alpha_2}{\sqrt{2}}|+\rangle_{e_1}|R(R-L)\rangle_{ab}\nonumber\\
&&-\frac{\beta_1\alpha_2}{\sqrt{2}}|+\rangle_{e_1}|L(R+L)\rangle_{ab}]\nonumber\\
&&\otimes(\gamma_1|a_1\rangle|-1\rangle_{e_2}+\delta_1|a_2\rangle|+1\rangle_{e_2})\nonumber\\
&&\otimes(\gamma_2|b_1\rangle+\delta_2|b_2\rangle).
\end{eqnarray}

Next, after another Hadamard operation is performed on NV$_1$, we
lead the wave packets of photon $c$ from the two spatial modes
$|c_1\rangle$ and $|c_2\rangle$ to $X$, $U$, NV$_1$, $U$, $X$, and CPBS in
sequence as shown in Fig.\ref{figure5}. The  quantum system
composed of NV$_1$, NV$_2$, and photons $a$, $b$, and $c$ is evolved
from $|\Phi_{abce_1e_2}\rangle_4\equiv
|\Phi_{abe_1e_2}\rangle_4\otimes|\phi_c\rangle_0$ to
\begin{eqnarray}                           
|\Phi_{abce_1e_2}\rangle_5 \!\!&=&\!\! [\alpha_1\beta_2|-1\rangle_{e_1}|RL\rangle_{ab}(\gamma_3|c_2\rangle+\delta_3|c_1\rangle)\nonumber\\
&&\!\!+\beta_1\beta_2|+1\rangle_{e_1}|LL\rangle_{ab}(\gamma_3|c_1\rangle+\delta_3|c_2\rangle)\nonumber\\
&&\!\!-\frac{\alpha_1\alpha_2}{\sqrt{2}}|-1\rangle_{e_1}|R(R-L)\rangle_{ab}(\gamma_3|c_2\rangle+\delta_3|c_1\rangle)\nonumber\\
&&\!\!-\frac{\beta_1\alpha_2}{\sqrt{2}}|-1\rangle_{e_1}|L(R+L)\rangle_{ab}(\gamma_3|c_2\rangle+\delta_3|c_1\rangle)]\nonumber\\
&&\!\!\otimes(\gamma_2|b_1\rangle+\delta_2|b_2\rangle)(\gamma_1|a_1\rangle|-1\rangle_{e_2}\nonumber\\
&&\!\!+\delta_1|a_2\rangle|+1\rangle_{e_2})(\alpha_3|R\rangle+\beta_3|L\rangle)_c.
\end{eqnarray}

Subsequently, the wave packets of photon $b$ from the two spatial
modes $|b_1\rangle$ and $|b_2\rangle$ are led to CPBS, NV$_1$, CPBS,
$X$, and $U$ (through the dash-dot-dotted line in Fig.\ref{figure5})
after a Hadamard operation is performed on NV$_1$, and then the
wavepackets of photon $c$ from the two spatial modes $|c_1\rangle$
and $|c_2\rangle$ are led to CPBSs, NV$_2$, CPBS, $X$, and $U$ in
sequence as shown in Fig.\ref{figure5}. The state of the system
$abce_1e_2$ is transformed from $|\Phi_{abce_1e_2}\rangle_5$ to
\begin{eqnarray}                           
|\Phi_{abce_1e_2}\rangle_6\!\!&=&\!\![\alpha_1\beta_2|+\rangle_{e_1}|RL\rangle_{ab}(\gamma_3|c_2\rangle+\delta_3|c_1\rangle)\nonumber\\
&&\!\!+\beta_1\beta_2|-\rangle_{e_1}|LL\rangle_{ab}(\gamma_3|c_1\rangle+\delta_3|c_2\rangle)\nonumber\\
&&\!\!-\frac{\alpha_1\alpha_2}{\sqrt{2}}|+\rangle_{e_1}|R(R-L)\rangle_{ab}(\gamma_3|c_2\rangle+\delta_3|c_1\rangle)\nonumber\\
&&\!\!+\frac{\beta_1\alpha_2}{\sqrt{2}}|-\rangle_{e_1}|L(R+L)\rangle_{ab}(\gamma_3|c_2\rangle+\delta_3|c_1\rangle)]\nonumber\\
&&\!\!\otimes(\gamma_2|b_1\rangle+\delta_2|b_2\rangle)(|\!-\!1\rangle_{e_2}\gamma_1\beta_3|a_1L\rangle_{ac}\nonumber\\
&&\!\!+|\!+\!1\rangle_{e_2}\delta_1\beta_3|a_2L\rangle_{ac}-|\!-\!1\rangle_{e_2}\gamma_1\alpha_3|a_1L\rangle_{ac}\nonumber\\
&&\!\!+|\!+\!1\rangle_{e_2}\delta_1\alpha_3|a_2R\rangle_{ac}).
\end{eqnarray}

After the Hadamard operations are performed on NV$_1$ and NV$_2$, we
lead the wave packets of photon $b$ from the two spatial modes
$|b_1\rangle$ and $|b_2\rangle$ to $H_P$, CPBS, NV$_1$, CPBS, $X$, and
$U$ again (through the dotted line in Fig.\ref{figure5}). And we lead
those from the two spatial modes $|c_1\rangle$ and $|c_2\rangle$ to
$H_P$, CPBS, NV$_2$, CPBS, $X$, and $U$ (through the dotted line in
Fig.\ref{figure5}), and then the state of the quantum system becomes
\begin{eqnarray}                           
|\Phi_{abce_1e_2}\rangle_7\!\!&=&\!\![\alpha_1\beta_2|-1\rangle_{e_1}|RL\rangle_{ab}(\gamma_3|c_2\rangle+\delta_3|c_1\rangle)\nonumber\\
&&\!\!+\beta_1\beta_2|+1\rangle_{e_1}|LL\rangle_{ab}(\gamma_3|c_1\rangle+\delta_3|c_2\rangle)\nonumber\\
&&\!\!+\alpha_1\alpha_2|-1\rangle_{e_1}|RR\rangle_{ab}(\gamma_3|c_2\rangle+\delta_3|c_1\rangle)\nonumber\\
&&\!\!+\beta_1\alpha_2|+1\rangle_{e_1}|LR\rangle_{ab}(\gamma_3|c_2\rangle+\delta_3|c_1\rangle)]\nonumber\\
&&\!\!\otimes[\gamma_1\beta_3|+\rangle_{e_2}|a_1L\rangle_{ac}+\delta_1\beta_3|-\rangle_{e_2}|a_2L\rangle_{ac}\nonumber\\
&&\!\!-\frac{\gamma_1\alpha_3}{\sqrt{2}}|+\rangle_{e_2}|a_1(R-L)\rangle_{ac}\nonumber\\
&&\!\!-\frac{\delta_1\alpha_3}{\sqrt{2}}|+\rangle_{e_2}|a_2(R+L)\rangle_{ac}]\nonumber\\
&&\!\!\otimes(\gamma_2|b_1\rangle+\delta_2|b_2\rangle).
\end{eqnarray}
After a Hadamard operation is performed on NV$_2$, we lead the
wave packets of photon $b$ from the two spatial modes $|b_1\rangle$
and $|b_2\rangle$ to CPBS, $X$, $U$, NV$_2$, $U$, $X$, and CPBS in sequence as
shown in Fig.\ref{figure5}. The state of the quantum system becomes
\begin{eqnarray}                           
|\Phi_{abce_1e_2}\rangle_8\!\!&=&\!\![\alpha_1\beta_2|-1\rangle_{e_1}|RL\rangle_{ab}(\gamma_3|c_2\rangle+\delta_3|c_1\rangle)\nonumber\\
&&\!\!+\beta_1\beta_2|+1\rangle_{e_1}|LL\rangle_{ab}(\gamma_3|c_1\rangle+\delta_3|c_2\rangle)\nonumber\\
&&\!\!+\alpha_1\alpha_2|-1\rangle_{e_1}|RR\rangle_{ab}(\gamma_3|c_2\rangle+\delta_3|c_1\rangle)\nonumber\\
&&\!\!+\beta_1\alpha_2|+1\rangle_{e_1}|LR\rangle_{ab}(\gamma_3|c_2\rangle+\delta_3|c_1\rangle)]\nonumber\\
&&\!\!\otimes[\gamma_1\beta_3|-1\rangle_{e_2}|a_1L\rangle_{ac}(\gamma_2|b_2\rangle+\delta_2|b_1\rangle)\nonumber\\
&&\!\!+\delta_1\beta_3|+1\rangle_{e_2}|a_2L\rangle_{ac}(\gamma_2|b_1\rangle+\delta_2|b_2\rangle)\nonumber\\
&&\!\!-\frac{\gamma_1\alpha_3}{\sqrt{2}}|\!-\!1\rangle_{e_2}|a_1(R\!-\!L)\rangle_{ac}(\gamma_2|b_2\rangle\!+\!\delta_2|b_1\rangle)\nonumber\\
&&\!\!-\frac{\delta_1\alpha_3}{\sqrt{2}}|\!-\!1\rangle_{e_2}|a_2(R\!+\!L)\rangle_{ac}(\gamma_2|b_2\rangle\!+\!\delta_2|b_1\rangle)].\nonumber\\
\end{eqnarray}

Next, after a Hadamard operation is performed on NV$_2$, we lead the
wavep ackets from the two spatial modes $|c_1\rangle$ and
$|c_2\rangle$ to CPBS, NV$_2$, CPBS, $X$, and $U$ (through the
dash-dot-dotted line in Fig.\ref{figure5}). The state of the quantum
system becomes
\begin{eqnarray}                           
|\Phi_{abce_1e_2}\rangle_9\!\!&=&\!\![|-1\rangle_{e_1}\alpha_1\beta_2|RL\rangle_{ab}(\gamma_3|c_2\rangle+\delta_3|c_1\rangle)\nonumber\\
&&\!\!+|+1\rangle_{e_1}\beta_1\beta_2|LL\rangle_{ab}(\gamma_3|c_1\rangle+\delta_3|c_2\rangle)\nonumber\\
&&\!\!+|-1\rangle_{e_1}\alpha_1\alpha_2|RR\rangle_{ab}(\gamma_3|c_2\rangle+\delta_3|c_1\rangle)\nonumber\\
&&\!\!+|+1\rangle_{e_1}\beta_1\alpha_2|LR\rangle_{ab}(\gamma_3|c_2\rangle+\delta_3|c_1\rangle)]\nonumber\\
&&\!\!\otimes[|+\rangle_{e_2}\gamma_1\beta_3|a_1L\rangle_{ac}(\gamma_2|b_2\rangle+\delta_2|b_1\rangle)\nonumber\\
&&\!\!+|-\rangle_{e_2}\delta_1\beta_3|a_2L\rangle_{ac}(\gamma_2|b_1\rangle+\delta_2|b_2\rangle)\nonumber\\
&&\!\!-\frac{\gamma_1\alpha_3}{\sqrt{2}}|+\rangle_{e_2}|a_1(R-L)\rangle_{ac}(\gamma_2|b_2\rangle+\delta_2|b_1\rangle)\nonumber\\
&&\!\!+\frac{\delta_1\alpha_3}{\sqrt{2}}|-\rangle_{e_2}|a_2(R+L)\rangle_{ac}(\gamma_2|b_2\rangle+\delta_2|b_1\rangle)].\nonumber\\
\end{eqnarray}
After another Hadamard operation is performed on NV$_2$ again, we
put the wave packets of photon $c$ from the two spatial modes
$|c_1\rangle$ and $|c_2\rangle$ into $H_P$, CPBS, NV$_2$, CPBS, $X$, $U$,
and CPBS (through the dotted line in Fig.\ref{figure5}), and the
state of the quantum system becomes
\begin{eqnarray}                           
|\Phi_{abce_1e_2}\rangle_{10}\!\!&=&\!\![|-1\rangle_{e_1}\alpha_1\beta_2|RL\rangle_{ab}(\gamma_3|c_2\rangle+\delta_3|c_1\rangle)\nonumber\\
&&\!\!+|+1\rangle_{e_1}\beta_1\beta_2|LL\rangle_{ab}(\gamma_3|c_1\rangle+\delta_3|c_2\rangle)\nonumber\\
&&\!\!+|-1\rangle_{e_1}\alpha_1\alpha_2|RR\rangle_{ab}(\gamma_3|c_2\rangle+\delta_3|c_1\rangle)\nonumber\\
&&\!\!+|+1\rangle_{e_1}\beta_1\alpha_2|LR\rangle_{ab}(\gamma_3|c_2\rangle+\delta_3|c_1\rangle)]\nonumber\\
&&\!\!\otimes[|-1\rangle_{e_2}\gamma_1\beta_3|a_1L\rangle_{ac}(\gamma_2|b_2\rangle+\delta_2|b_1\rangle)\nonumber\\
&&\!\!+|+1\rangle_{e_2}\delta_1\beta_3|a_2L\rangle_{ac}(\gamma_2|b_1\rangle+\delta_2|b_2\rangle)\nonumber\\
&&\!\!+|-1\rangle_{e_2}\gamma_1\alpha_3|a_1R\rangle_{ac}(\gamma_2|b_2\rangle+\delta_2|b_1\rangle)\nonumber\\
&&\!\!+|+1\rangle_{e_2}\delta_1\alpha_3|a_2R\rangle_{ac}(\gamma_2|b_2\rangle+\delta_2|b_1\rangle)].\nonumber\\
\end{eqnarray}

At last, we perform a Hadamard operation  on each of NV$_1$ and
NV$_2$, and then the spatial-mode bit-flip operations are performed on photons $b$
and $c$. By measuring the states of NV$_1$ and NV$_2$ with the
orthogonal basis $\{|-1\rangle, |+1\rangle\}$, the outcome of the
hybrid hyper-Toffoli gate on a three-photon system can be obtained
by performing the conditional operations on photon $a$. If NV$_1$ is
projected into the state $|+1\rangle_{e_1}$, a polarization
operation $|L\rangle_a\rightarrow -|L\rangle_a$ is performed on
photon $a$. If NV$_2$ is projected into the state
$|+1\rangle_{e_2}$, a spatial-mode operation $|a_2\rangle\rightarrow
-|a_2\rangle$ is performed on photon $a$. In this way, the state of
the three-photon system $abc$ becomes
\begin{eqnarray}                           
|\Phi_{abc}\rangle\!\!&=&\!\![(\alpha_1\beta_2|RL\rangle_{ab}+\alpha_1\alpha_2|RR\rangle_{ab}+\beta_1\alpha_2|LR\rangle_{ab})\nonumber\\
&&\!\!(\gamma_3|c_1\rangle+\delta_3|c_2\rangle)+\beta_1\beta_2|LL\rangle_{ab}(\gamma_3|c_2\rangle+\delta_3|c_1\rangle)]\nonumber\\
&&\!\![(\gamma_1\beta_3|a_1L\rangle_{ac}+\gamma_1\alpha_3|a_1R\rangle_{ac}+\delta_1\alpha_3|a_2R\rangle_{ac})\nonumber\\
&&\!\!(\gamma_2|b_1\rangle+\delta_2|b_2\rangle)+\delta_1\beta_3|a_2L\rangle_{ac}(\gamma_2|b_2\rangle+\delta_2|b_1\rangle)].\nonumber\\
\end{eqnarray}
This is the result of the hybrid hyper-Toffoli gate operating on the
three-photon system $abc$.

\section{Discussion and Summary} \label{sec5}

An NV center in diamond is an appropriate dipole emitter in cavity
QED to obtain the high-fidelity reflection-transmission property in
the Purcell regime, with its long spin coherence time ($\sim$ms)
\cite{NV3,NV5} and nanosecond manipulation time \cite{NV6}. When a
diamond NV center is coupled to a micro- or nano- cavity,  the
spontaneous emission of dipole emitter into the zero-phonon line can
be greatly enhanced, and the interaction of the NV center and the
photon is also enhanced \cite{CNV2,CNV5}. There are many
experimental demonstrations of diamond NV centers coupled to micro-
or nano- resonators with either a strong-coupling strength
\cite{CNV3} or a weak-coupling one \cite{CNV4}. In 2012, Faraon
\emph{et al.} \cite{CNV5} demonstrated experimentally that the
zero-phonon transition rate of an NV center is greatly enhanced
($\sim70$) by coupling to a photonic crystal resonator ($Q\sim3000$)
fabricated in a monocrystalline diamond with the coupling strength
as a few GHz, and they pointed out that the photonic crystal
platform with a quality factor of $Q\sim10^5$ can operate at the
onset of strong-coupling regime.

\begin{figure}[htb]                    
\centering
\includegraphics[width=7.2 cm]{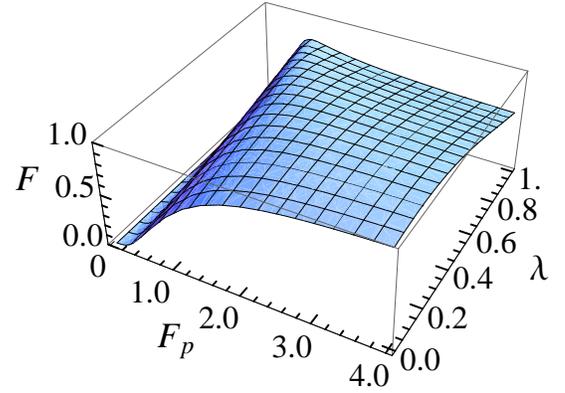}
\caption{(Color online)  Fidelity ($F$) of a hybrid
spatial-polarization hyper-CNOT gate vs the Purcell factor $F_P$ and
the cavity decay rate $\lambda$.} \label{figure6}
\end{figure}

In the double-sided cavity-NV-center system, the two waveguides are
simultaneously coupled to the cavity resonator mode with the
coupling constants $\eta_1$ and $\eta_2$, respectively. As the
backscattering is low in the waveguide, the asymmetry of the two
coupling constants is mainly caused by cavity intrinsic loss
$\kappa$ \cite{FOS}. In experiment, the difference of the two coupling
constants $\Delta\eta\sim0.2\eta$ has been demonstrated, which
yields approximately the same fidelity for both transmission and
reflection directions \cite{FOS}. The reflection and transmission
coefficients of a double-sided cavity-NV-center system are dominated
by the Purcell factor $F_P$ and the cavity decay rate
$\lambda=\kappa/\eta$. In the resonant condition
($\omega_c=\omega_k=\omega$), the transmission and reflection rule
for a handedness circularly polarized light can be described as
\begin{eqnarray} \label{eq16}       
\begin{split}
|R^\uparrow, -1\rangle     \;\; \rightarrow \;\;&  |rL^\downarrow+tR^\uparrow, -1\rangle, \\
|L^\downarrow, -1\rangle   \;\; \rightarrow \;\;&  |rR^\uparrow+tL^\downarrow, -1\rangle, \\
|R^\downarrow, +1\rangle   \;\; \rightarrow \;\;&  |rL^\uparrow+tR^\downarrow, +1\rangle, \\
|L^\uparrow, +1\rangle     \;\; \rightarrow \;\;&  |rR^\downarrow+tL^\uparrow, +1\rangle, \\
|R^\downarrow, -1\rangle   \;\; \rightarrow \;\;&  |t_0R^\downarrow+r_0L^\uparrow, -1\rangle, \\
|L^\uparrow, -1\rangle     \;\; \rightarrow \;\;&  |t_0L^\uparrow+r_0R^\downarrow, -1\rangle, \\
|R^\uparrow, +1\rangle     \;\; \rightarrow \;\;&  |t_0R^\uparrow+r_0L^\downarrow, +1\rangle, \\
|L^\downarrow, +1\rangle   \;\; \rightarrow \;\;&
|t_0L^\downarrow+r_0R^\uparrow, +1\rangle.
\end{split}
\end{eqnarray}

The fidelity of the photonic quantum logic gate can be calculated by
$F=\overline{|\langle\psi_f|\psi\rangle|^2}$, where $|\psi\rangle$
is the ideal finial state of a quantum logical gate, and
$|\psi_f\rangle$ is the finial state of a quantum system by
considering experimental factors ($\alpha_i, \beta_i, \gamma_i,
\delta_i \in[0,1]$). The fidelity of our hybrid photonic hyper-CNOT
gate is shown in Fig.\ref{figure6}, which is decreased with a small
Purcell factor or a large cavity intrinsic loss. In Fig.\ref{figure6}, the
fidelity of the hybrid photonic hyper-CNOT gate is higher with a small
Purcell factor when the cavity intrinsic loss becomes larger, which
corresponds to the regime $|r|\simeq|t_0|$. That is, the fidelity of
the hybrid photonic hyper-CNOT gate is higher when the
reflection-transmission properties of the uncoupled cavity and the
coupled cavity are symmetrical. In the case $|r|=|t_0|$, the
relation of the Purcell factor and the cavity decay rate is
$F_P=(1-\frac{\lambda^2}{4})/\lambda$.


\begin{figure}[htb]                    
\centering
\includegraphics[width=7.2 cm]{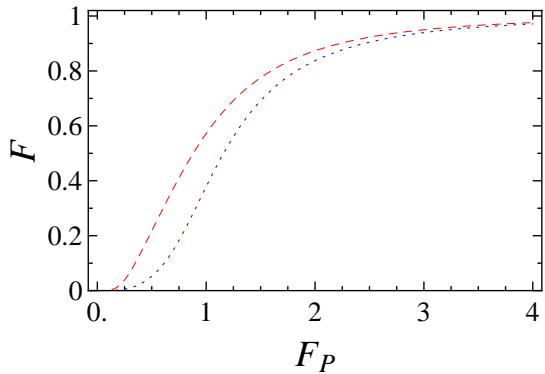}
\caption{(Color online)  Fidelity ($F$) of a hybrid spatial-polarization
hyper-CNOT gate on a two-photon system (red dashed line) and that of
two identical polarization CNOT gates on a four-photon system (blue
dotted line) vs the Purcell factor $F_P$. Here the cavity decay rate
is chosen as $\lambda=0.1$, and the construction of the polarization
CNOT gate is the same as that of the polarization part of the
hyper-CNOT gate in Ref. \cite{QD4}.} \label{figure7}
\end{figure}

\begin{figure}
\centering
\includegraphics[width=7.2 cm]{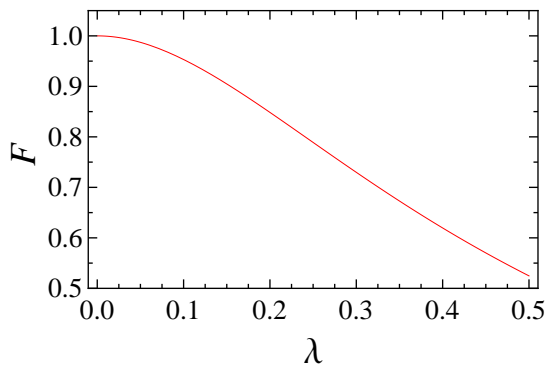}
\caption{(Color online)  Fidelity ($F$) of a hybrid spatial-polarization
hyper-Toffoli gate vs the cavity decay rate $\lambda$ in the case $|r|=|t_0|$
[$F_P=(1-\frac{\lambda^2}{4})/\lambda$], which is equal to the
one for two polarization Toffoli gates on a six-photon system. } \label{figure8}
\end{figure}

The probability of recovering an incident photon after operation is
increased with a large cavity decay rate $\eta$ \cite{FOS}. O'Shea
\emph{et al.}  \cite{FOS} noted that the maximal fidelity of the
operation with cavity QED is achieved at the regime where the
coupling strength $g$ is smaller than the cavity decay rate $\eta$
($F_P>1$) rather than the strong-coupling regime, and the maximal
fidelity is obtained at the point $\lambda\simeq0.1$ in their
experiment. In the case $\lambda=0.1$, the fidelity of our  hybrid
photonic hyper-CNOT gate on a two-photon system and that of the two
identical CNOT gates on a four-photon system in one DOF are shown in
Fig.\ref{figure7}. It shows that the fidelity of our hybrid photonic
hyper-CNOT gate is higher than the one for the two CNOT gates in one
DOF in the weak-coupling regime with a small Purcell factor. In the
case $|r|=|t_0|$, the fidelity of the hybrid hyperparallel photonic
logic gate is equal to the one for the two identical photonic logic
gates in one DOF (e.g., the fidelity of the hybrid hyper-Toffoli
gate shown in Fig.\ref{figure8}). In the case $\lambda=0.1$ and
$F_P=9.875$, both the fidelity of the hybrid photonic hyper-CNOT
gate and that of the two identical CNOT gates in one DOF are
$F=99.7\%$. That is, the hybrid hyperparallel photonic logic gate
can decrease the effect of environment noise in the asymmetric
condition of the double-sided cavity-NV-center system in the
weak-coupling regime with a small Purcell factor.

The reflection property of one-sided dipole-cavity protocols is
fragile because the reflectance for the uncoupled cavity and the
coupled cavity should be balanced to get a high fidelity, while the
reflection-transmission property of double-sided dipole-cavity
systems is robust and flexible with the large reflectance and
transmittance difference between the uncoupled cavity and the
coupled cavity \cite{QD,QD4}. Moreover, a double-sided dipole-cavity
system has two spatial modes, so it is very convenient to use this
DIT to investigate the robust and flexible quantum information
processing based on the polarization and spatial-mode DOFs of photon
systems.

Both CONT and Toffoli gates are parts of the set of universal quantum logic gates,
and they can form universal quantum computing with the assistance of
single-qubit rotation gates \cite{QC}. Both our hybrid
polarization-spatial hyper-CNOT gate and hyper-Toffoli gate can form
universal hyperparallel photonic quantum computing assisted by the
rotations on a single photon in two DOFs, which is useful in
the quantum information protocols with multi-qubit
systems in several DOFs. For example, hyperentanglement is useful in
quantum communication protocols for increasing the channel capacity \cite{HQC},
resorting to the entanglement in several DOFs of photon systems \cite{heper1}.
With hyperparallel quantum gates, the generation and complete analysis
of hyperentangled states can be achieved in a relatively simpler way,
compared with the protocols with several cascaded quantum entangling gates
\cite{multiqubit2,multiqubit3,multiqubit4}. Besides, some quantum
information processes can be implemented with less resources based on several
DOFs of photon systems, resorting to the hyperparallel quantum gates. For example,
in the preparation of four-qubit cluster states, only a hyper-CNOT gate operation (photons
interact with electron spins four times) and a wave plate are
required in the protocol with two photons in two DOFs \cite{multiqubit6}, while three
CNOT gate operations (photons interact with electron spins six
times) are required in the protocol with four photons in one DOF.

In summary, we have presented the DIT of a double-sided
cavity-NV-center system, which is still obvious in the weak-coupling
regime. The reflection-transmission property of circularly polarized
light interacting with a double-sided cavity-NV-center system can be
used for photon-photon interaction in quantum information processing
based on both the polarization and spatial-mode DOFs. With the DIT
of double-sided cavity-NV-center systems, we have proposed a hybrid
photonic hyper-CNOT gate and a hybrid photonic hyper-Toffoli gate
for hyperparallel photonic quantum computation. A hyperparallel
hybrid quantum logic gate on a quantum system in both the
polarization and spatial-mode DOFs is equal to the two identical
quantum gates operating on that in one DOF simultaneously, and it can
depress the resource consumption, photonic dissipation, and
asymmetric environment noise of the double-sided cavity-NV-center system
in the weak-coupling regime with a small Purcell factor. Besides,
these hyperparallel quantum logic gates are useful for the quantum
information protocols with multi-qubit systems in several DOFs,
especially the generation and analysis of hyperentangled states
\cite{multiqubit2,multiqubit3,multiqubit4}.

The double-sided cavity QED can be used for quantum information
processing even in a bad cavity regime (the Purcell regime)
 \cite{QD,QD4,RT}, and it is suitable to investigate the robust and
flexible quantum information processing based on both the
polarization and spatial-mode DOFs \cite{multiqubit2,multiqubit3,multiqubit4}, according to its
reflection-transmission optical property. Besides the quantum
computation with two DOFs of a photon as two qubits
\cite{twoDOF,twoDOF1}, double-sided cavity QED can also be used for
quantum information processing with two DOFs by using a photon as a
qudit. Moreover, the multiqubit logic gate based on one DOF can be
simplified with less photon resources by resorting to two DOFs of
photon systems \cite{multiqubit}.

\section*{ACKNOWLEDGMENTS}

This work is supported by the National Natural Science Foundation of China under Grants Nos. 11174039 and 11474026, and NECT-11-0031.

\end{document}